\documentclass[article,aps,12pt]{revtex4}
\usepackage{graphicx}
\begin{document}

\title{
   Quasi Two-Dimensional Electron--Hole Systems \\
   in a Perpendicular Magnetic Field}

\author{
   John J. Quinn$^1$, 
   Arkadiusz Wojs$^{1,2}$, and 
   Kyung-Soo Yi$^{1,3}$}

\affiliation{
   $^1$University of Tennessee, Knoxville, Tennessee 37996, U.S.A.\\
   $^2$Wroclaw University of Technology, 50-370 Wroclaw, Poland\\
   $^3$Pusan National University, Busan 609-735, Korea}

\begin{abstract}
The electronic properties of quasi two-dimensional multicomponent systems are investigated in the 
presence of a perpendicular magnetic field.
The effects of the presence of a few valence band holes on the properties of quantum Hall systems 
are examined by analyzing the results of exact numerical diagonalization of small electron-hole 
systems confined to the spherical surface.
The novel type of elementary excitations, the angular momenta, binding energies, interaction 
pseudopotentials, and effects on the photoluminescence spectrum are presented.
\end{abstract}
\maketitle
{RUNNING TITLE: \sl Quasi Two-Dimensional Electron--Hole Systems}

\section*{TABLE OF CONTENTS}

   I. INTRODUCTION

  II. EXACT DIAGONALIZATION OF SMALL SYSTEMS

 III. EXCITONS AND EXCITONIC COMPLEXES

  IV. PSEUDOPOTENTIALS

   V. GENERALIZED COMPOSITE FERMION MODEL

  VI. CONDENSED STATES OF CHARGED EXCITONS

 VII. OTHER CHARGE CONFIGURATIONS

VIII. PHOTOLUMINESCENCE

\hspace{0.5cm}A. $N$-Electron--One-Hole Systems

\hspace{0.5cm}B. Photoluminescence Selection Rules at High Magnetic Fields

\hspace{0.5cm}C. Singlet and Triplet Charged Excitons at Low Magnetic Fields

\hspace{0.5cm}D. Skyrmion Excitons

  IX. SUMMARY

\clearpage

\section{Introduction}

In semiconducting inversion layers and heterostructures, electrons can be confined to a very thin
layer in one direction, but they are able to move freely in the plane perpendicular to the direction
of confinement.
These effectively two-dimensional electron systems (2DES) have been of great interest for more than
three decades \cite{ando,bastard}.
Because the electron concentration can be varied over a wide range of values within a single sample,
these 2D systems are ideal testing grounds for studying many body interactions.

The application of a large dc magnetic field perpendicular to the 2D layer results in some notable
novel physics.
The single electron states coalesce into highly degenerate Landau levels whose energy is given 
by $\varepsilon_n=\hbar \omega_c (n+1/2)$ where $\omega_c$ is equal to the electron cyclotron
frequency $eB/mc$, and $n$ is a non-negative integer.
Each Landau level can hold $N_\phi=B\mathcal{A}/\phi_0$ electrons of a given spin,
where $\mathcal{A}$ is the area of the sample and $\phi_0=hc/e$ is the quantum of flux.
The integral quantum Hall effect \cite{iqhe} occurs when $N$ electrons exactly fill an integral
number of Landau levels resulting in an integral value of the filling factor $\nu=N/N_\phi$.
In that case the ground state is incompressible because an infinitesimal decrease in the area
$\mathcal{A}$, which decreases $N_\phi$, requires a finite energy $\hbar \omega_c$ to promote an
electron to the next Landau level.

\begin{figure}
\resizebox{13.0cm}{15.0cm}{\includegraphics{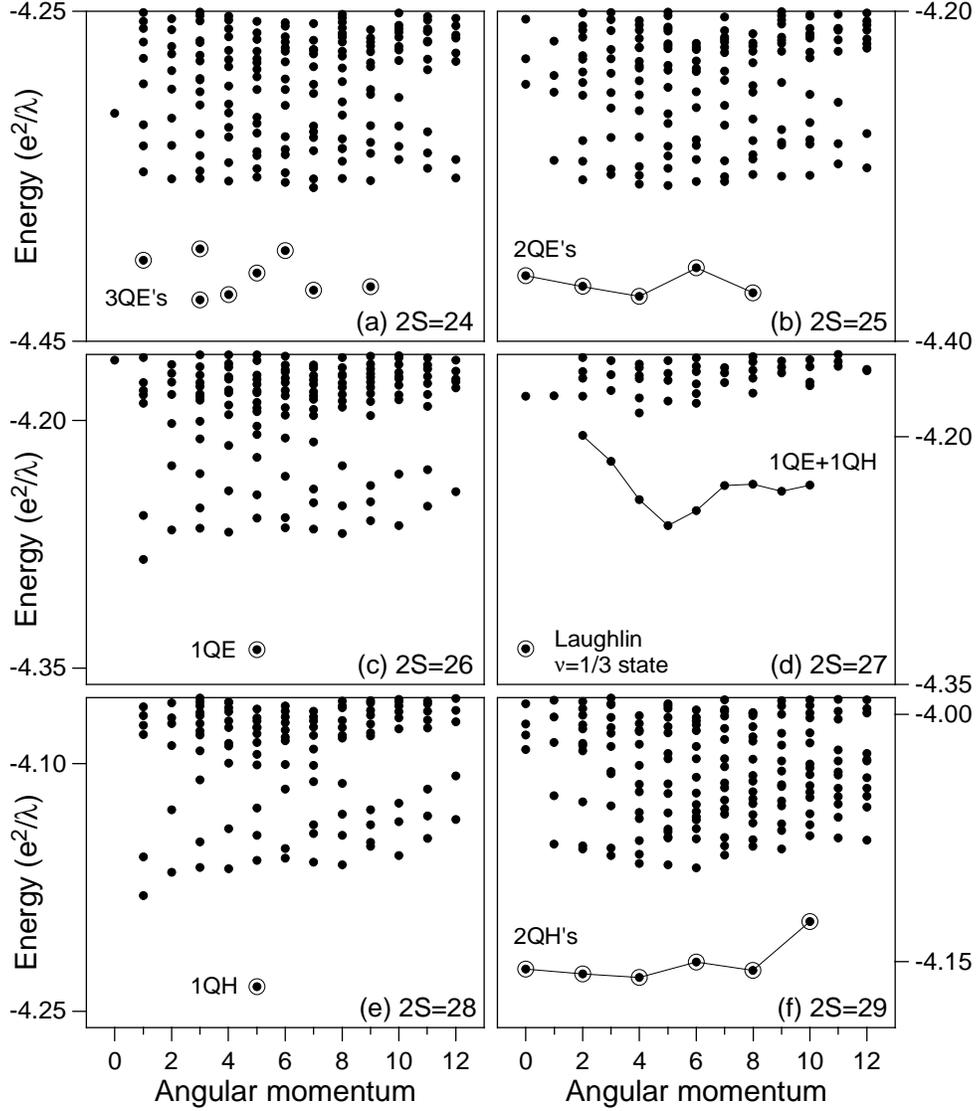}}
\caption{\label{fig1}
   The energy spectra of 10 electrons in the lowest Landau level calculated on a Haldane sphere with
   $2Q$ between 24 and 29.
   The open circles and solid lines mark the lowest energy bands with the fewest composite Fermion 
   quasiparticles \cite{acta phys}.}
\end{figure}

It is more difficult to understand why fractional quantum Hall states \cite{tsui,laughlin} are
incompressible \cite{jpcm,pseudo,varenna}.
When $N_\phi$ is larger than $N$, no gap occurs in the absence of electron--electron interactions.
At very high values of the applied magnetic field, there is only one relevant energy scale in the 
problem, the Coulomb scale $e^2/\lambda$, where $\lambda=(\hbar c/eB)^{1/2}$ is the magnetic 
length.
In that case standard many body perturbation theory is inapplicable.
With remarkable insight, Laughlin proposed a ground state wavefunction and the form of the
elementary excitations that contained the essential correlations responsible for the energy
gap \cite{laughlin}.
Exact diagonalization of the interaction Hamiltonian within the Hilbert subspace of the lowest 
Landau level is a very good approximation at large values of $B$ 
(where $\hbar \omega_c \gg e^2/\lambda$).
Although it can be carried out only for relatively small systems, it gives beautiful confirmation 
of Laughlin's picture of the essential correlations. 
The numerical diagonalization studies are usually carried out on a spherical surface of radius $R$
on which the electrons are confined \cite{numerical,fano}.
A magnetic monopole of strength $2Q\phi_0$, where $2Q$ is an integer, is located at the origin.
It produces a radial magnetic field $\vec{B}=(2Q\phi_0/4\pi R^2)\,\hat r$.
The single particle eigenstates are denoted by $\left|Q,l,m\right>$ and are called monopole
harmonics \cite{fano}.
They are eigenfunctions of ${\hat l}^2$, $l_z$, and the single particle Hamiltonian $H$, with 
eigenvalues $l(l+1)$, $m$, and $(\hbar \omega_c/2Q)[l(l+1)-Q^2]$ respectively \cite{wu}.
Because the energy eigenvalue must be positive, the allowed values of $l$ are given by $l_n=Q+n$,
where $n=0, 1, 2, \dots$.
The lowest Landau level or angular momentum shell has $l_0=Q$ and the allowed values of $m$
have $|m|\leq Q$.
An $N$-electron eigenfunction can be written
\begin{equation}
\left|m_1,m_2,\dots,m_N\right>=
C_{m_N}^\dagger \dots C_{m_2}^\dagger C_{m_1}^\dagger \left|0\right>.
\end{equation}
Here $C_m^\dagger$ creates an electron in state $\left|l_0, m\right>$.
There are $G_{NQ}=N_\phi!/N!/(N_\phi-N)!$ of $N$-electron states in the Hilbert space 
of the lowest Landau level (with degeneracy $N_\phi=2Q+1$).
The numerical problem is to diagonalize the interaction Hamiltonian in this $G_{NQ}$ 
dimensional space.
Typical numerical results are shown in Fig.~\ref{fig1} for a system of 10 electrons with value of
$2Q$ between 24 and 29 \cite{acta phys}.

It is well-known in quantum field theory that the presence of a Chern--Simons gauge field, 
obtained by attaching to each electron an infinitesimally thin flux tube carrying magnetic 
flux $\alpha \phi_0$, has 
no effect on the classical equations of motion \cite{wilczek}.
The reason for this is that the Chern--Simons magnetic field is given by 
$\vec{b} (\vec{r} )=\alpha\phi_0 \sum_i \delta(\vec{r}-\vec{r}_i ) \hat z$,
where $\vec{r}_i$ is the position of the $i^{th}$ electron and $\hat z$ is a unit vector parallel to
the applied $\vec{B}$ field.
Because two electrons cannot occupy the same position, a given electron never senses the 
Chern--Simons magnetic
field due to any other electron.
However, it does experience the vector potential
\begin{equation}
\vec{a} (\vec{r} )=\alpha \phi_0 \int d^2 r_1 \frac{\hat z \times 
(\vec{r}-\vec{r}_1 )}{|\vec{r}-\vec{r}_1|^2}
\Psi^\dagger (\vec{r}_1) \Psi(\vec{r}_1),
\label{gaugefield}
\end{equation}
where $\Psi^\dagger (\vec{r}_1)\Psi(\vec{r}_1)$ is the density operator for the electron fluid.
Adding $\vec{a} (\vec{r} )$ to $\vec{a} (\vec{r} )=\frac{1}{2} B(\hat z \times \vec{r} )$, the 
vector
potential of the dc magnetic field, gives a very complex many body Hamiltonian for a quantum
mechanical system.
It is usually treated by starting with a mean field approximation in which 
$\Psi^\dagger (r)\Psi(r)$ is replaced by its ground state expectation value $n_S$, 
the uniform equilibrium density, in Eq.~(\ref{gaugefield}).
Then, the mean field Hamiltonian becomes a sum of one-particle Hamiltonians in the presence of the 
mean
magnetic field $B^*=B+\alpha\phi_0 n_S$.
Despite the lack of a small parameter, fluctuations beyond the mean field have been treated by
standard many body perturbation theory with qualitative success \cite{halperin,lopez}.

Jain introduced the idea of a composite Fermion (CF) to represent an electron with an attached flux
tube which carries an even number $\alpha$ ($=2p$) of flux quanta \cite{cf}.
In the mean field approximation the CF filling factor $\nu^*$ is given by 
${\nu^*}^{-1}=\nu^{-1}-\alpha$, i.e. the number
of flux quanta per electron of the dc field less the CF flux per electron.
When $\nu^*$ is equal to an integer $n=\pm1, \pm2, \dots$, then 
$\nu=n (1+\alpha n)^{-1}$ generates
(for $\alpha=2$) quantum Hall states at $\nu=1/3, 2/5, 3/7, \dots$ and 
$\nu=1, 2/3, 3/5, \dots$.
These are the most pronounced FQH states observed.

In the spherical geometry one can introduce an effective monopole strength seen by one
CF \cite{chen}.
It is given by $2Q^*=2Q-\alpha (N-1)$ since the $\alpha$ flux quanta attached to every other CF
must be subtracted from the original monopole strength $2Q$.
Then $|Q^*|=l_0^*$ plays the role of the angular momentum of the lowest CF shell just as $Q=l_0$ 
was the angular momentum of the lowest electron shell.
When $2Q$ is equal to an odd integer $(1+\alpha)$ times $(N-1)$, the CF shell $l_0^*$ is completely
filled, and an $L=0$ incompressible Laughlin state at filling factor $\nu=(1+\alpha)^{-1}$
results.
When $2|Q^*|+1$ is smaller or larger than $N$, quasielectrons (QE's) or quasiholes (QH's) 
appear in the shell with $l_{\rm QE}=l_0^*+1$ or $l_{\rm QH}=l_0^*$, respectively.
The low energy sector of the energy spectrum consists of the states with the minimum number 
of quasiparticle excitations required by the value of $2Q^*$ and $N$.
The first excited band of states will contain one additional QE--QH pair.
The total angular momentum of these states in the lowest energy sector can be predicted by addition 
of the angular momenta ($l_{\rm QH}$ or $l_{\rm QE}$) of
the $n_{\rm QH}$ or $n_{\rm QE}$ quasiparticles (QE's or QH's) 
treated as identical Fermions.
In Tab.~\ref{tab1} we demonstrated how these allowed $L$ values are found 
for a 10-electron system with $2Q$ in the range $29 \ge 2Q \ge 21$. 
By comparing with numerical results \cite{acta phys} presented in Fig.~\ref{fig1}, 
we readily observe that the total angular momentum multiplets appearing in the lowest energy 
sector are always correctly predicted by this simple mean field Chern--Simons picture.

It is quite surprising that this mean field Chern--Simons picture works so well.
Fluctuations beyond the mean field interact via both Coulomb and Chern--Simons gauge interactions.
The mean field Chern--Simons picture introduces a new energy scale $\hbar \omega_c^*$ proportional 
to the effective 
magnetic field $B^*$, in addition to the Coulomb scale $e^2/\lambda$.
For large values of the applied magnetic field, this new energy scale is very large compared with 
the Coulomb scale, but it is totally irrelevant 
to the determination of the low energy spectrum.

\begin{table}
\caption{\label{tab1}
   The effective CF monopole strength $2Q^*$, 
   the CF quasiparticle numbers $n_{\rm QH}$ and $n_{\rm QE}$, 
   the quasiparticle angular momenta $l_{\rm QE}$ and $l_{\rm QH}$, and 
   the angular momenta $L$ of the lowest lying band of multiplets 
   for a 10-electron system at $2Q$ between 29 and 21.}
\begin{tabular}{|l||c|c|c|c|c|c|c|c|c|} \hline
$2Q$&29&28&27&26&25&24&23&22&21\\ \hline \hline
$2Q^*$&11&10&9&8&7&6&5&4&3\\  \hline
$n_{\rm QH}$&2&1&0&0&0&0&0&0&0\\  \hline
$n_{\rm QE}$&0&0&0&1&2&3&4&5&6\\   \hline
$l_{\rm QH}$&5.5&5&4.5&4&3.5&3&2.5&2&1.5\\  \hline
$l_{\rm QE}$&6.5&6&5.5&5&4.5&4&3.5&3&2.5\\   \hline
$L$  & 10,8,6,4,2,0 &5 &0 &5 &$8,6,4,2,0$ 
&$9,7,6,5,4,3^2,1$ & $8,6,5,4^2,2^2,0$
&5,3,1 &0 \\ \hline
\end{tabular}
\end{table}

The reason for the success of the mean field Chern--Simons picture has been discussed in detail in 
terms of the
behavior of the electron pseudopotential $V(L')$, the interaction energy of a pair of electrons as a
function of the total angular momentum $L'$ of the pair 
\cite{fermiliquid,pseudo,jpcm,lee,no411,prb61}.
If $V(L')$ is of the form $V_{\rm H}(L')=A+BL'(L'+1)$, which we refer to as a ``harmonic''
pseudopotential,
then every angular momentum multiplet which has the same value of the total angular momentum $L$ has
the same energy.
In other words, the interactions, which couple only states with the same total angular momentum,
introduce no correlations.
Any linear combination of eigenstates with the same total angular momentum has the same energy.
If $V(L')$ increases more quickly with increasing $L'$ than $V_{\rm H}(L')$, then Laughlin
correlations, maximum possible avoidance of pair amplitudes for the largest values of $L'$, give the 
lowest energy states.
We call such pseudopotentials {\em superharmonic}.
For harmonic and subharmonic pseudopotentials, Laughlin correlations do not occur.

Properties of 2D electron systems are most frequently studied by transport measurements, or by
optical measurements \cite{ando,bastard}.
The latter include infrared absorption and reflection, inelastic  light scattering, and
photoluminescence.
In all of these processes, valence band holes are either created or destroyed.
Therefore, it is important to include the valence band holes in the theoretical investigations of
the many body correlations.

In this review we investigate how the properties of quantum Hall systems are affected by the
presence of a few valence band holes, either confined to the same 2D surface as the 2DES, or 
separated from it by a finite distance $d$ \cite{jpcm}.
When $d$ is equal to zero, the strengths of the electron--electron and electron--hole interactions
are equal in magnitude.
Then a ``hidden symmetry'' results from the fact that the commutator of the Hamiltonian with the
operator $d^\dagger(0)$, which creates an exciton of momentum zero, satisfies the relation
\cite{lerner,dzyubenko,macdonald}
\begin{equation}
[\hat H, d^\dagger(0)]=E_X(0) d^\dagger (0).
\label{commutator}
\end{equation}
Here $d^\dagger(\vec{k})$ creates an exciton of wavevector $\vec{k}$ and $E_X(0)$ is the 
binding energy of a single exciton.
The operator $d^\dagger (0)$ can be written
\begin{equation}
d^\dagger (0)=N_\phi^{-1/2} \sum_{k'} c_{k'}^\dagger b_{-k'}^\dagger,
\end{equation}
where $N_\phi=2Q+1$ and $c_{k'}^\dagger$ (or $b_{k'}^\dagger$) creates an electron (or hole) in the 
lowest Landau 
level with wavenumber in the
$y$-direction (in the Landau gauge) given by $k'$.
Because of Eq.~(\ref{commutator}), if $\left|\phi\right>$ is an eigenfunction of $\hat H$ with 
eigenvalue
$E_\phi$,
then $d^\dagger (0)\left|\phi\right>$ is an eigenfunction of $\hat H$ with energy $E_\phi+E_X$.
In other words, neutral excitons affect the energy of the electron system only by adding the energy
of the bare exciton to the total electron energy.
The neutral exciton is essentially uncoupled from the electron system, and states containing only 
$N_X$ neutral excitons and $N_e$ free unbound electrons are called ``multiplicative states''.
Although in some circumstances they can be the lowest energy states, multiplicative states are not
necessarily the ground states \cite{prl70,prb51-10880}.

This paper is organized as follows:
In Section II, we present results of exact numerical diagonalization of small electron--hole systems 
confined to the same spherical surface under the assumption that $\hbar \omega_c \gg e^2/\lambda$.
In Section III, states containing neutral excitons (X), negatively charged excitons ($X^-$), and 
larger excitonic complexes ($X_n^-$=an electron bound to $n$ neutral excitons) are found.
From the analysis of their energy spectra we can determine the binding energy of the excitonic
complexes.
In Section IV, the pseudopotential describing the energy of interaction $V_{\rm AB}(L')$ of a pair
(A and B can represent any of these charged Fermions: $e$, $X^-$, $X_2^-$, \dots) as a 
function of the angular momentum of the pair is extracted from the numerical results.
In Section V, a generalized composite particle model for a multicomponent quantum liquid is 
introduced.
Because the pseudopotentials $V_{\rm AB}(L')$ are ``superharmonic'', i.e., increase faster than 
$L'(L'+1)$with increasing $L'$, Laughlin correlations among the constituents can occur.
In Section VI, for a system containing only negatively charged excitons ($X^-$), Laughlin 
condensed states of these $X^-$ Fermions are found at particular values of the $X^-$ filling factor 
$\nu_{X^-}$.
However, other charge configurations can also exist.
For example, one $X^-$ can break up into an electron and a neutral X.
Knowledge of the pseudopotentials $V_{\rm AB}(L')$ allows us to evaluate the low lying energy
spectra for each charge configuration.
This is presented in Section VII.

The main connection of our theoretical work with experiment is through the analysis of
photoluminescence (PL).
In this experimental technique, commonly used in studying the 2DES, a small number of conduction
electron--valence hole pairs are excited in a quantum well containing $10^{11}$ to $10^{12}$ 
electrons/${\rm cm}^2$.
Because $N_e \gg N_h$, the separation between valence band holes is very large, and it
is only necessary to study the radiative electron--hole recombination of a single hole interacting 
with the conduction electrons.
The luminescence energy depends on the energy of the $N_e$-electron--one-hole initial state 
and that of the $N_e-1$ electron final state.
The energy spectra of a system of nine electrons and one valence hole as a function of the 
separation
$d$ between the layer containing the 2DES and that containing the hole is presented in Section
VIII A.
Several different values of the monopole strength $2Q$ are considered, and the interpretation of
prominent bands of states is given.
In Section VIII B, selection rules for the radiative recombination of the hole with an electron are
discussed, and a few simple examples are given.
In these sections it was assumed that the electrons and hole were confined to 2D planes, and that 
$\hbar \omega_c \gg e^2/\lambda$, so that only the states in the lowest Landau level had to be
considered.
For more realistic experimental systems, the ratio of $\hbar \omega_c$ to $e^2/\lambda$ need not be
large.
Then, the Hilbert subspace in which the diagonalization is performed must contain an admixture of
higher Landau levels.
In addition, the effect of finite quantum well widths must be included.
This is done for very small systems (two or three electrons and one hole) in Section VIII C.
The total electron spin must be specified, because the negatively charged excitons can form in
singlet ($J_e=0$) or triplet ($J_e=1$) spin states.
The diagonalization gives us the eigenfunctions as well as the eigenvalues of all the
eigenstates.
The former can be used to study the oscillator strength for PL.
In Section VIII D, we consider the spin flip excitations of a Landau level at filling factor of
unity.
These excitations are spin waves, and their dispersion relation and interactions can be obtained
from the numerical results. 
When the filling factor $\nu$ differs from unity due to the presence of one hole ($h$) in the
$\nu=1$ level (or one spin reversed electron ($e_{\rm R}$) in the higher spin level) additional 
pairs of
holes and spin reversed electrons are spontaneously formed and bound to the hole (or to the spin
reversed electron).
The number of such $h$--$e_{\rm R}$ pairs depends on the Zeeman energy, and the resulting 
``spin textures'' are called antiskyrmions (or skyrmions) \cite{skyrmions,cote,paredes,wqskyrmion}.
They are very much analogous to the $X_n^-$ excitonic complexes.
When a valence band hole is introduced into an electron system at a filling factor of unity, it
should create and bind $h-e_{\rm R}$ pairs to form ``skyrmion excitons.''
It seems quite likely that skyrmion excitons play an important role in PL near filling factor
$\nu=1$. 
The final section is a summary of results on electron--hole systems in 2D.
%The novel types of elementary excitations, their angular momenta, binding energies, interaction
%pseudopotentials, and effect on the PL spectrum are summarized.

\section{Exact Diagonalization of Small Systems}

In the presence of a strong applied magnetic field, neutral excitons $X$ and spin-polarized charged
excitonic complexes $X_k^-$ can occur.
These $X_k^-$ complexes consist of $k$ neutral excitons bound to an electron and must be
distinguished from spin-unpolarized ones (e.g. the singlet $X^-$) that occur at lower magnetic
fields but unbind when $\hbar \omega_c \gg e^2/\lambda$.
The $X_k^-$ complexes are long lived Fermions whose spectra display Landau level structure.
Here we perform exact numerical diagonalization (within the subspace of the lowest Landau level) for
small systems containing $N_e$ electrons and $N_h$ holes ($N_e>N_h$).

The numerical diagonalization for the electron--hole system is a simple extension of that for
electrons alone.
One can select states that have a definite value of $M$, the $z$-component of the total angular
momentum.
The holes are Fermions, but distinguishable from the electrons.
Therefore, we can construct states in which the $N_e$ electrons occupy states with distinct
values of $m$, the $z$-component of the single electron angular momentum, and the $N_h$ holes
occupy states with distinct values of $m'$, the $z$-component of the hole angular momentum.
Because the electrons and holes are distinguishable, the values of $m$ and $m'$ selected from the
range between $-l$ and $l$ need not be different for the two types of Fermions.
When the interaction Hamiltonian is diagonalized within the subspace of the lowest Landau level, the
eigenvalues $E(L)$ are obtained.

\section{Excitons and Excitonic Complexes}

\begin{figure}
\resizebox{8.0cm}{8.0cm}{\includegraphics{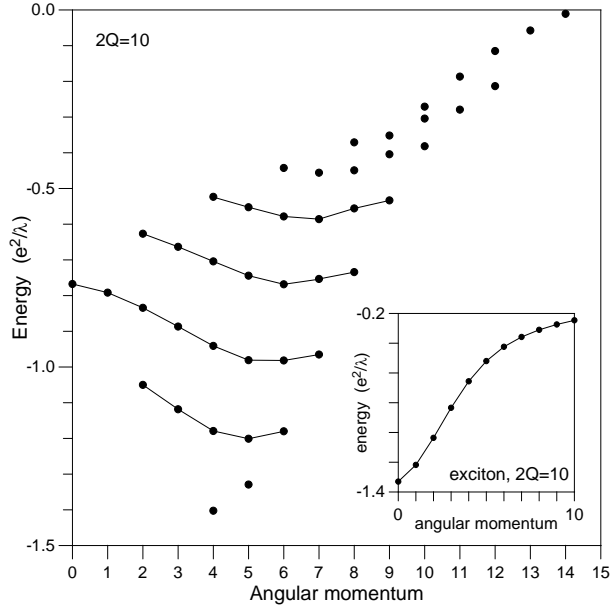}}
\caption{\label{fig2}
  Energy spectrum of two electrons and one hole at $2Q=10$.
  Inset: energy spectrum of an electron-hole pair \cite{prb60-2}.}
\end{figure}

In Fig.~\ref{fig2}, we show the energy spectrum (in magnetic units) of a system of two electrons and 
one hole
at $2Q=10$ as a function of the total angular momentum $L$ \cite{prb60-2}.
The lowest energy state at $L=Q$ is the multiplicative state with one neutral exciton in its $l_X=0$
ground state and one electron of angular momentum $l_e=Q$.
Only one state of lower energy occurs in the spectrum.
It appears at $L=Q-1$ and corresponds to the only bound state of the negatively charged exciton
$X^-$.
The value of the $X^-$ angular momentum, $l_{X^-}=Q-1$, can be understood by noticing that the
lowest energy single particle configuration of the two electrons and one hole is the 
``compact droplet'', in which the two electrons have $z$-component of angular momentum $m=Q$ and
$m=Q-1$, and the hole has $m=-Q$ giving $M=Q-1$.

As marked with lines in Fig.~\ref{fig2}, unbound states above the multiplicative state form bands, 
which arise from the $e$--$h$ interaction and are separated by gaps associated with the 
characteristic 
excitation energies of an $e$--$h$ pair. (The $e$--$h$ pseudopotential, i.e., the energy spectrum
of an exciton, is shown in the inset). 
These bands are rather well approximated by the expectation values of the total ($e$--$e$ and 
$e$--$h$) interaction energy, calculated in the eigenstates of the $e$--$h$ interaction alone 
without $e$--$e$ interaction.

\begin{figure}
\resizebox{9.0cm}{8.0cm}{\includegraphics{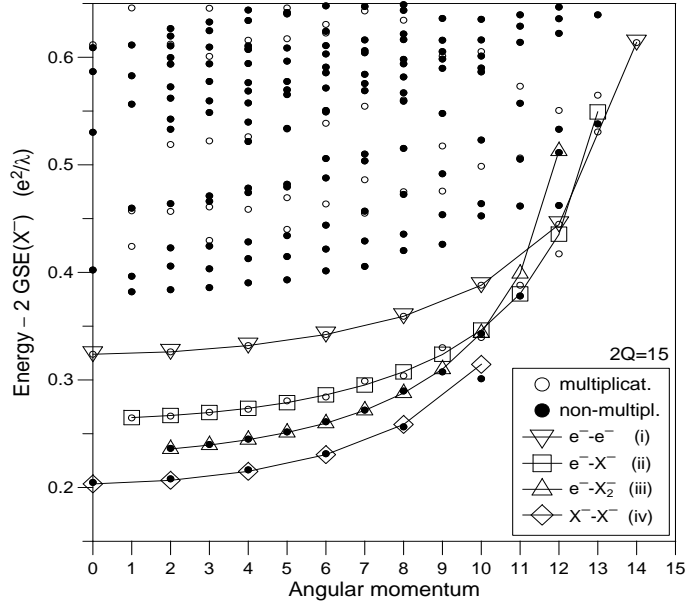}}
\caption{\label{fig3}
   Energy spectrum of four electrons and two holes at $2Q=15$.
   Open circles: multiplicative states; solid circles: 
    non-multiplicative states; triangles, 
   squares, and diamonds: approximate pseudopotentials 
   \cite{prb60-2}.}
\end{figure}

In Fig.~\ref{fig3}, we display the energy spectrum obtained by numerical 
diagonalization of the Coulomb interaction of a system of four electrons 
and two holes at $2Q=15$ \cite{prb60-2}.
The states marked by open and solid circles are multiplicative 
(containing one or more decoupled $X$'s) and non-multiplicative states, 
respectively.
For $L\le10$ there are four rather well defined low lying bands.
Two of them begin at $L=0$.
The lower of these consists of two $X^-$ ions interacting through 
a pseudopotential $V_{X^--X^-}(L')$.
The upper band consists of states containing two decoupled $X$'s 
plus two electrons interacting through $V_{e^--e^-}(L')$.
The band that begins at $L=1$ consists of one $X$ plus an $X^-$ 
and an electron interacting through $V_{e^--X^-}(L')$, while the band 
which starts at $L=2$ consists of an $X_2^-$ interacting with a free 
electron.

Knowing that the angular momentum of an electron is $l_{e^-}=Q$, 
we can see that $l_{X_k^-}=Q-k$, and that decoupled excitons do not 
carry angular momentum ($l_X=0$).
For a pair of identical Fermions of angular momentum $l$ the allowed 
values of the pair angular momentum are $L'=2l-j$, where $j$ is an 
odd integer.
For a pair of distinguishable particles with angular momentum $l_A$ 
and $l_B$, the total angular momentum satisfies $|l_A-l_B|\leq L' 
\leq l_A+l_B$.
The states containing two free electrons and two decoupled neutral 
excitons fit exactly the pseudopotential for a pair of electrons 
at $2Q=15$; the maximum pair angular momentum is $L'^{MAX}=14$ as 
expected.
By comparing this band of states with the band containing two 
$X^-$'s, we can obtain the binding energy of the neutral exciton 
to the electron to form the $X^-$.
The other binding energy, that of a neutral exciton to an $X^-$ 
to form an $X_2^-$ can be obtained in a similar way.

We define $\varepsilon_0$ as the binding energy of a neutral exciton, 
$\varepsilon_1$ as the binding energy of an $X$ to an electron to form 
an $X^-$, and $\varepsilon_k$ as the binding energy of an $X$ to an 
$X_{k-1}^-$ to form an $X_k^-$.
An estimate of these binding energies (in magnetic energy units 
$e^2/\lambda$ where $\lambda$ is the magnetic length) as a function 
of $2Q$ are given in Tab.~\ref{tab2} \cite{prb60-1}.
We note clearly that 
$\varepsilon_0>\varepsilon_1>\varepsilon_2>\varepsilon_3$.

\begin{table}
\caption{\label{tab2}
Binding energies $\varepsilon_0$, $\varepsilon_1$, $\varepsilon_2$, 
$\varepsilon_3$ of $X, X^-,X_2^-$, and $X_3^-$, respectively, 
in units of $e^2/\lambda$ \cite{prb60-1}.}
\begin{tabular}{|l||c|c|c|c|} \hline
$2Q$&$\varepsilon_0$&$\varepsilon_1$&$\varepsilon_2$&$\varepsilon_3$\\ 
\hline \hline
10&1.3295043&0.0728357&0.0411069&0.0252268\\  \hline
15&1.3045679&0.0677108&0.0395282&0.0262927\\  \hline
20&1.2919313&0.0647886&0.0381324&0.0260328\\   \hline \hline
\end{tabular}
\end{table}

\section{Pseudopotentials}

In Fig.~\ref{fig3}, the band of states containing two $X^-$'s terminates 
at $L'=10$. 
Since the $X^-$'s are Fermions, one would have expected a state at 
$L'^{MAX}=2l_{X^-}-1=12$.
This state is missing in Fig.~\ref{fig3}.
We surmise that the state with $L'=L'^{MAX}$ does not occur because 
of the finite size of the $X^-$.
Large pair angular momentum corresponds to the small average separation, 
and two $X^-$'s in the state with $L'^{MAX}$ would be too close to one 
another for the bound $X^-$'s to remain stable.
We can think of this as a ``hard core'' repulsion for $L'=L'^{MAX}$.
Effectively, the corresponding pseudopotential parameter, 
$V_{X^--X^-}(L'^{MAX})$ is infinite.
In a similar way, $V_{e^--X^-}(L')$ is infinite for $L'=L'^{MAX}=14$, 
and $V_{e^--X_2^-}(L')$ is infinite for $L'=L'^{MAX}=13$.

Once the maximum allowed angular momenta for all four pairings $AB$ 
are established, all four bands in Fig.~\ref{fig3} can be roughly 
approximated by the pseudopotentials of a pair of point charges with
angular momentum $l_A$ and $l_B$, shifted by the binding energies 
of appropriate composite particles.
For example, the $X^--X^-$ band is approximated by the $e^--e^-$ pseudopotential for $l=l_{X^-}=Q-1$
plus twice the $X^-$ energy.
The agreement is demonstrated in Fig.~\ref{fig3}, where the squares, diamonds, and two kinds of 
triangles
approximate the four bands in the four-electron--two-hole spectrum.
The fit of the diamonds to the actual $X^--X^-$ spectrum is quite good 
for $L'<10$.
The fit of the $e^--X^-$ squares to the open circle multiplicative 
states is reasonably good for $L'<12$, and the $e^--X_2^-$ triangles 
fit their solid circle non-multiplicative states rather well for $L'<11$.
At sufficiently large separation (low $L'$), the repulsion between 
ions is weaker than their binding and the bands for distinct charge 
configurations do not overlap.

There are two important differences between the pseudopotentials 
$V_{AB}(L')$ involving composite particles and those involving 
point particles.
The main difference is the hard core discussed above.
If we define the relative angular momentum $\mathcal{R}=l_A+l_B-L'$ 
for a pair of particles with angular momentum $l_A$ and $l_B$, then 
the minimum allowed relative angular momentum (which avoids the hard 
core) is found to be given by
\begin{equation}
\mathcal{R}_{AB}^{min}=2{\rm min}(k_A,k_B)+1,
\end{equation}
where $A=X_{k_A}^-$ and $B=X_{k_B}^-$.
The other difference involves polarization of the composite particle.
A dipole moment is induced on the composite particle by the electric 
field of the charged particles with which it is interacting.
By associating an ``ionic polarizability'' with the excitonic ion 
$X_k^-$, the polarization contribution to the pseudopotential can 
easily be estimated.
When a number of charges interact with a given composite particle, 
the polarization effect is reduced from that caused by a single charge, 
because the total electric field at the position of the excitonic 
ion is the vector sum of contributions from all the other charges, 
and there is usually some cancellation.
We will ignore this effect in the present work and simply use the 
pseudopotential $V_{AB}(L')$ obtained from Fig.~\ref{fig3} to 
describe the effective interaction.

\section{Generalized Composite Fermion Model}

When the pseudopotential describing the interaction of a pair of identical Fermions as a function of
the pair angular momentum is ``superharmonic'', the low energy states are those with
Laughlin--Jastrow correlations.
By this we mean that these low energy states avoid to the maximum possible extent having pairs with
pair angular momentum $L'=2l-1$, its maximum value.
If Chern--Simons flux is added adiabatically to the interacting Fermions, Laughlin correlations
automatically results.
The Laughlin correlations allow us to describe the interacting Fermions in terms of an effective
angular momentum $l^*=l-p(N-1)$ where $N$ is the number of identical Fermions, and $p$ is an
integer.
The value of $l^*$ can be thought of in terms of a composite Fermion (CF) transformation in which 
the monopole strength $2Q$ is replaced by $2Q^*=2Q-2p(N-1)$, i.e., the original monopole strength
seen by one electron minus the Chern--Simons flux on every other electron (to which $2p$ 
Chern--Simons flux quanta
oppositely oriented to the applied magnetic field have been added).
This ``effective'' angular momentum $l^*$ results in the electrons having as a largest possible pair
angular momentum of $L'^*=2l^*-1=2l-1-2p(N-1)$.
By forbidding states with the largest allowed electron pair angular momenta, we do not allow pair
states with the smallest spatial size and the largest repulsion.

We know from our numerical calculations that we can obtain states with neutral excitons, electrons,
charged triplet excitons $X^-$ and higher spin polarized excitonic complexes $X_k^-$.
The charged particles are all Fermions with Landau levels.
Clearly, the low energy sector of the energy spectrum can contain different types of charged
Fermions: electrons, $X^-$, $X_2^-, \dots$.
It is straightforward to construct a generalized CF picture for a multicomponent Fermion plasma.
Let's define the effective monopole strength $2Q_a^*$ seen by a CF of type $a$ as 
\begin{equation}
2Q_a^*=2Q-\sum_b (m_{ab}-\delta_{ab})(N_b-\delta_{ab}).
\end{equation}
What we have done here is to attach to all type $a$ Fermions $(m_{aa}-1)$ flux quanta that couple
only to the charges on all other type $a$ Fermions and $m_{ab}$ flux quanta sensed only by charges
on the type $b$ Fermions.
This is a straightforward generalization of what we did in making the CF transformation for a one
component (electron) plasma.
The coefficients $m_{ab}$ are the powers that occur in the generalized Laughlin wavefunction,
$\prod_{\left<i,j\right>} (z_i^{(a)}-z_j^{(b)})^{m_{ab}}$ where $z_i^{(a)}$ is the complex 
coordinate of the
$i^{th}$ Fermion of type $a$ and the product is over all pairs $\left<i,j\right>$.
For different multicomponent systems generalized Laughlin incompressible states are expected to
occur when
(i) all the hard-core pseudopotentials are avoided and
(ii) each type of CF's (i.e., CF$_a$'s, CF$_b$'s, \dots) completely fills an integral
number of their angular momentum shells.
In other cases, low lying multiplets are expected to contain different kinds of CF quasiparticles
(QE$_a$'s, QE$_b$'s, \dots, or QH$_a$'s, QH$_b$'s, \dots) of the incompressible
generalized Laughlin states.

\section{Condensed States of Charged Excitons}

Consider for a moment a system containing 12 electrons and six holes on a Haldane spherical
surface at monopole strength $2Q=17$.
The charge configuration with the largest binding energy is that containing six $X^-$ charged
excitons.
We will refer to it as (i); its total binding energy $\varepsilon_i$ is equal to 
$6(\varepsilon_0+\varepsilon_1)$.
If we make a CF transformation on this system of $N_{X^-}=6$ negatively charged excitons, we obtain
$2Q_{X^-}^*=2Q-2(N_{X^-}-1)=7$.
The angular momentum of the $X^-$ is given by $l_{X^-}=Q-1=15/2$ and that of the CF $X^-$ by
$l_{X^-}^*=Q_{X^-}^*-1=5/2$.
This means that the six CF $X^-$'s completely fill the $l_{X^-}^*=5/2$ shell
giving a Laughlin $L=0$ incompressible state at $\nu_{X^-}=1/3$.
Note that $2l=\nu^{-1} (N-1)$ holds for the quantum liquid of $X^-$'s just as it did in the case of
electrons.

Although the $X^-$ particles have relatively long lifetimes for radiative recombination of an
electron--hole pair, it seems unlikely that the Laughlin condensed state of negatively charged
excitons can be observed by the standard experimental techniques used in the case of condensed
states of an electron liquid.
One point worth noting is that the generalized CF picture of a multicomponent  plasma can be thought
of in terms of fictitious CF fluxes and CF charges that have different ``colors''.
For example, electrons could have a red Chern--Simons charge and $X^-$'s a green charge.
Then $(m_{ee}-1)$ red and $m_{eX^-}$ green Chern--Simons fluxes would be attached to each electron,
while $(m_{X^-X^-}-1)$ green and $m_{X^-e}$ red Chern--Simons fluxes would be attached to each 
$X^-$.

\section{Other Charge Configurations}

For the 12-electron--6-hole system, other charge configurations besides the six $X^-$'s can
occur as excited states.
Among these are (ii) $e^-+5X^-+X$ with total binding energy 
$\varepsilon_{ii}=6\varepsilon_0+5\varepsilon_1$, and (iii) $e^-+4X^-+X_2^-$ with total energy
$\varepsilon_{iii}=6\varepsilon_0+5\varepsilon_1+\varepsilon_2$.
The total energy of any state depends on the interaction energy of the constituent charged particles
as well as the binding energy.

\begin{figure}
\resizebox{13.0cm}{6.5cm}{\includegraphics{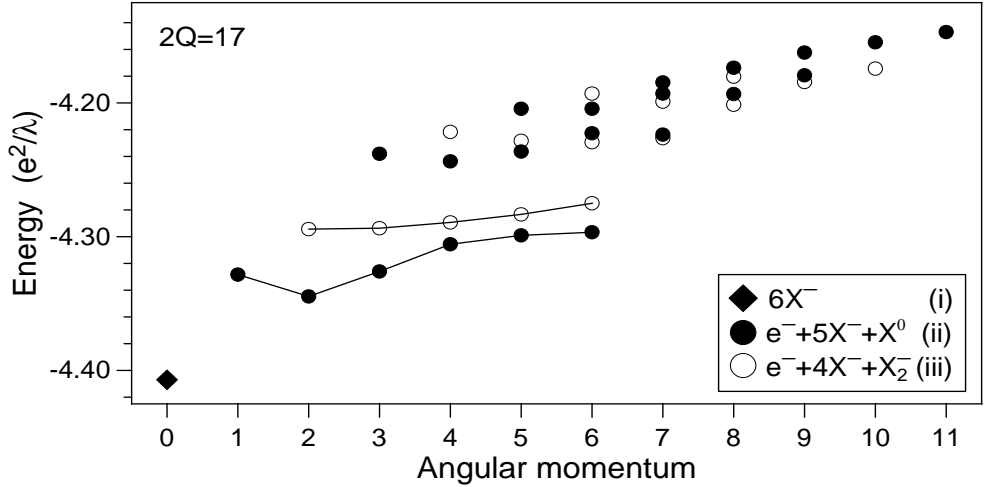}}
\caption{\label{fig4}
   Low energy spectra of different charge configurations of the $12e+6h$ system on a Haldane sphere
   at $2Q=17$: $6X^-$ (diamonds), $e^-+5X^-+X$ (solid circles), 
   and $e^-+4X^-+X_2^-$ (open circles) \cite{prb60-1}.}
\end{figure}

The system of eighteen particles (12 electrons and 6 holes) at $2Q=17$ is too large for us to
diagonalize in terms of the electrons and holes and their interactions.
However, we can obtain a reasonable approximation to the low lying energy spectrum by considering
the different charge configurations denoted by (i) through (iii) each of which contains only six
charged Fermions.
We make use of our knowledge of the binding energies, angular momenta, and pseudopotentials
$V_{AB}(L')$ where $A$ and $B$ can be $e^-$, $X^-$, or $X_2^-$.
The results of this simpler numerical calculation are presented in Fig.~\ref{fig4} \cite{prb60-1}.
There is only one low lying state of the six $X^-$ configuration, the $L=0$ Laughlin 
$\nu_{X^-}=1/3$ state.
There are two bands of states in each of the charge configurations (ii) and (iii).

The results presented in Fig.~\ref{fig4} can be understood from the generalized CF model.
The CF predictions are: (i) For the system of $N_{X^-}=6$, we take $m_{X^-X^-}=3$ and obtain the
Laughlin $\nu_{X^-}=1/3$ state as discussed earlier.
Because of the hard core of the $X^--X^-$ pseudopotential, this is the only state of this charge
configuration.
(ii) For the $e^-+5X^-+X$ configuration, we can take $m_{X^-X^-}=3$ and $m_{eX^-}=1, 2$, 
or 3.
For $m_{eX^-}=1$, we obtain $L=1, 2, 3^2, 4^2, 5^3, 6^3, 7^3, 8^2, 9^2, 10$, and 11.
For $m_{eX^-}=2$, we obtain $L=1$,  2, 3, 4, 5, and 6.
For $m_{eX^-}=3$, we obtain $L=1$.
(iii) For the grouping $e^-+4X^-+X_2^-$, we set $m_{X^-X^-}=3$, $m_{eX_2^-}=1$,
$m_{X^-X_2^-}=3$, and $m_{eX^-}=1, 2$, or 3.
For $m_{eX^-}=1$, we obtain the multiplets $L=2, 3, 4^2, 5^2, 6^3, 7^2, 8^2, 9$, and 10.
For $m_{eX^-}=2$, we obtain $L=2, 3, 4, 5$, and 6, and for $m_{eX^-}=3$, we obtain
$L=2$.  
In the groupings (ii) and (iii) the sets of multiplets obtained for higher values of 
$m_{eX^-}$ are subsets of those obtained for lower values of $m_{eX^-}$.
We would expect them to form lower energy bands since they avoid additional small values of
$\mathcal{R}_{eX^-}$.
However, note that the (ii) and (iii) states predicted for $m_{eX^-}=3$ (at $L=1$ and 2,
respectively) do not form separate bands in Fig.~\ref{fig4}.
This is because the $V_{eX^-}$ pseudopotential increases more slowly than linearly as a function of
$L'(L'+1)$ in the vicinity of $\mathcal{R}_{eX^-}=3$.
In such case the CF picture fails \cite{pseudo}.

The agreement of our CF predictions with the data in Fig.~\ref{fig4} is really quite
remarkable and strongly indicates that our multicomponent CF picture is correct.
We were indeed able to confirm predicted Jastrow type correlations in the low lying states by
calculating their coefficients of fractional parentage \cite{shell,pseudo}.
We have also verified the CF predictions for other systems that we were able to treat numerically.
If exponents $m_{ab}$ are chosen correctly, the CF picture works well in all cases.

\section{Photoluminescence}

In photoluminescence (PL) experiments the absorption of light creates a small number of
electron--hole pairs in a quantum well that already contains a concentration of conduction
electrons.
The valence band holes interact with the system of electrons.
The ultimate $e$--$h$ recombination results in an emitted photon whose energy is equal to the energy
difference between the initial state of $N$ electrons and one valence hole and the final state
containing $N-1$ electrons.
The initial state depends on the $e$--$h$ interaction, which in turn depends on the reparation 
between
the plane containing the $N$ electrons and the plane of the valence hole.
Such separation, which occurs, for example, in asymmetrically doped quantum wells, breaks the hidden
symmetry.
This allows for a rich PL spectrum which can be used as a probe of the low lying $e$--$h$ states 
even
at very high magnetic field, where only states within the lowest Landau level are of importance.
Because the number of valence band holes excited optically is so small, it is sufficient to study
the eigenstates of a single hole interacting with the $N$ electrons in the quantum well.
\begin{figure}
\resizebox{15.0cm}{17.5cm}{\includegraphics{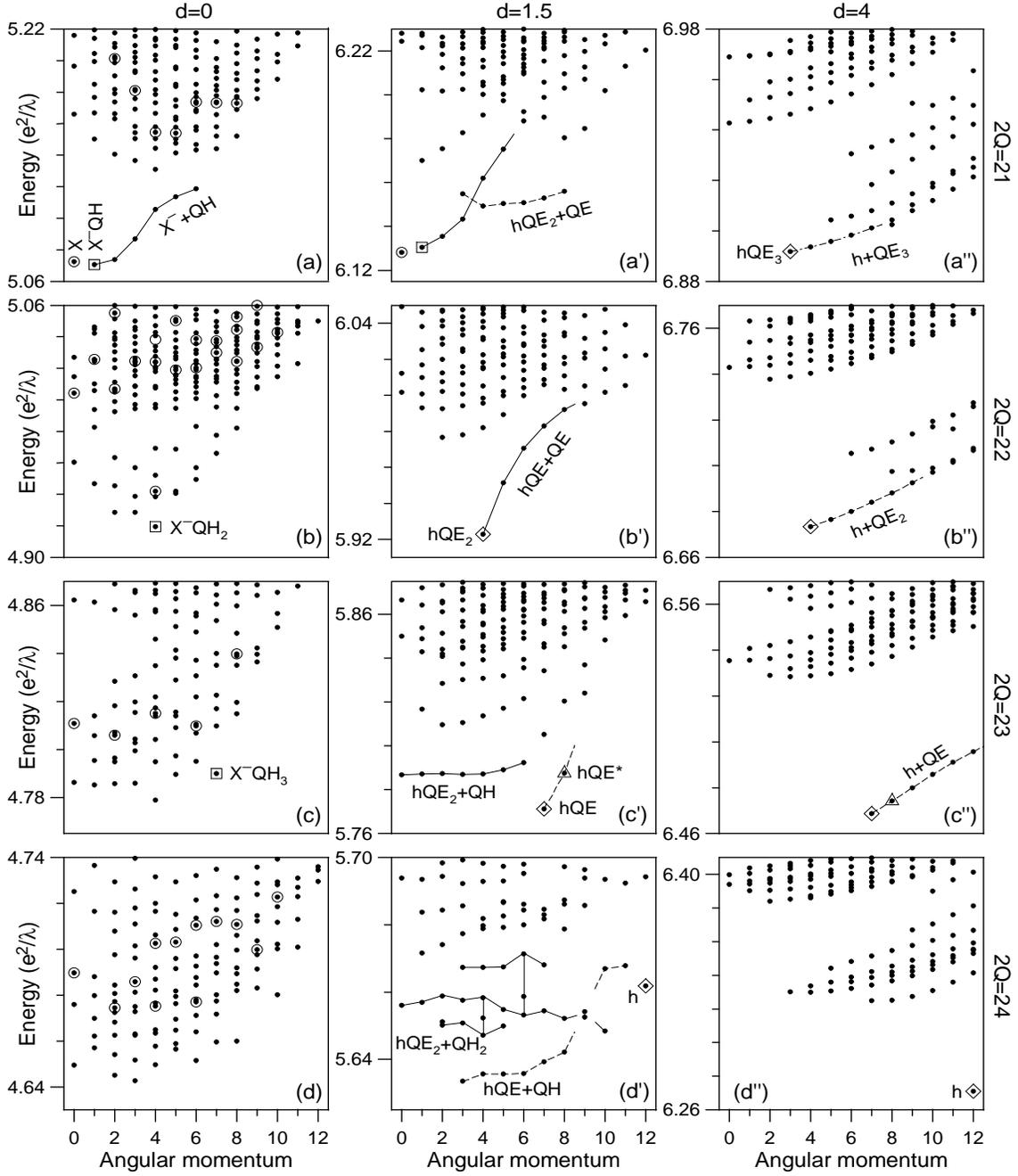}}
\caption{\label{fig5}
   Energy spectra of the nine-electron--one-hole system for the monopole strength 
   $2Q=21$, 22, 23, 24 (from top to bottom), and for the interplane separation 
   $d=0$, 1.5, 4 (from left to fight).
   Lines and open symbols mark the low lying states containing different bound 
   excitonic complexes \cite{pe11}.}
\end{figure}

\subsection{$N$-Electron--One-Hole Systems}

We start by considering the eigenstates of a system consisting of $N$ electrons confined to a plane
$z=0$ and interacting with one another and with a valence hole confined to a plane $z=d$, where $d$
is measured in units of the magnetic length $\lambda$.
The cyclotron energy $\hbar \omega_c$ is assumed to be much larger than the Coulomb energy
$e^2/\lambda$, so that only the lowest Landau level enters our calculation.
In Fig.~\ref{fig5} we present the energy spectra for a system of nine electrons and one valence band 
hole at
different values of the separation between electron and hole layers, and at different values of the
monopole strength $2Q$ \cite{pe11}.
For $d \ll 1$ we have strong coupling between the electrons and the hole.
Neutral ($X$) and charged triplet excitons ($X^-$) are found.
The multiplicative states at $d=0$ are shown as solid dots surrounded by a small circle.
Non-multiplicative states at $d=0$ can have an $X_t^-$ exciton interacting with the remaining $N-2$
electrons.
For $d \gg 1$ the valence hole interacts very weakly with the $N$-electron system, and the spectra
can be described in terms of the eigenstates of the $N$-electron system multiplied by the
eigenfunction of the hole with angular momentum $\hat L=\hat L_e+\hat l_h$.
For intermediate values of $d$ ($d \simeq 2$) the $e$--$h$ interaction is not a weak perturbation on
the electronic eigenstates, but it is not strong enough to bind a full electron to form an exciton.

For $d=0$, $X$ and $X^-$ bound states occur.
Due to the ``hidden symmetry'', the multiplicative states containing an $X$ have the same spectrum 
as
the eight electron system shifted by the $X$ binding energy.
The CF model \cite{cf,chen} tells us that the effective monopole strength seen by one CF in a system
of $N'=N-1=8$ electrons near $\nu=1/3$ is $2Q^*=2Q-2(N'-1)$.
$Q^*$ plays the role of the angular momentum of the lowest CF shell (Landau level), therefore
$Q^*=3.5$, 4, 4.5, and 5 for the multiplicative states in frames (a), (b), (c), and (d) of 
Fig.~\ref{fig5},
respectively.
Since the lowest shell can accommodate $2Q^*+1$ CF's, it is exactly filled in Fig.~\ref{fig5}(a), 
but there
are 1, 2, and 3 excess CF's for Fig.~\ref{fig5}(b), (c), and (d), respectively.
The excess CF's go into the next shell as Laughlin QE's with $l_{\rm QE}=Q^*+1$, giving one QE with
$l_{\rm QE}=4$(b), two QE's each with $l_{\rm QE}=4.5$(c), and three QE's each with 
$l_{\rm QE}=5$(d).
The angular momentum of the lowest band of multiplicative states are obtained by the addition of the
angular momentum of the QE excitations, remembering that they are identical Fermions.
These states are shown as points surrounded by a small circle in all frames for $d=0$.
In the absence of QE--QE interactions (i.e., for mean field CF's) all the states in the lowest CF
band of each spectrum would degenerate, but QE--QE interactions remove this degeneracy.
Higher energy multiplicative states that appear in the figure contain additional QE--QH pairs.

For the non-multiplicative states we have one $X^-$ and $N_e=N-2$ remaining electrons.
The generalized CF picture \cite{prb60-1} allows us to predict the lowest energy band in the
spectrum in the following way.
The effective monopole strength seen by the electrons is $2Q^*=2Q-2(N_e-1)-2 N_{X^-}$, while
that seen by the $X^-$ is $2Q_{X^-}^*=2Q-2N_e$.
Here, we have attached to each Fermion (electron and $X^-$) two fictitious flux quanta and used the
mean field approximation to describe the effective monopole strength seen by each particle (note
that a CF does not see its own flux).
The angular momentum of the lowest CF electron shell is $l_0^*=Q_e^*$, while that of the CF $X^-$
shell is $l_{X^-}^*=Q_{X^-}^*-1$ \cite{pb256,prb60-2}.
For the system with $N_e=7$ and $N_{X^-}=1$ at $2Q=21$, 22, 23, and 24, the generalized CF picture
leads to: one QH with $l_{\rm QH}=3.5$ and one $X^-$ with $l_{X^-}^*=2.5$, giving a band at 
$1 \leq L \leq 6$ for Fig.~\ref{fig5}(a);
two QH's with $l_{\rm QH}=4$ and one $X^-$ with $l_{X^-}^*=3$ giving 
$L=0\oplus 1\oplus 2^3 \oplus 3^3 \oplus 4^4 \oplus 5^3 \oplus 6^3 \oplus 7^2 \oplus 8^2 \oplus 9
\oplus 10$ for Fig.~\ref{fig5}(b);
three QH's with $l_{\rm QH}=4.5$ and one $X^-$ with $l_{X^-}^*=3.5$ for Fig.~\ref{fig5}(c);
and four QH's with $l_{\rm QH}=5$ plus one $X^-$ with $l_{X^-}^*=4$ for Fig.~\ref{fig5}(d).
In the figure, we have restricted the values of $L$ and of $E$, so that not all the states are
shown.

For $d \gg 1$, the electron--hole interaction is a weak perturbation on the energies obtained for
the $N$-electron system \cite{prl70,prb50,prb51}.
The numerical results can be understood by adding the angular momentum of the hole, $l_h=Q$, to the
electron angular momentum obtained from the simple CF model.
The predictions are: for $2Q=21$ there are three QE's each with $l_{\rm QE}=3.5$ and the hole has 
$l_h=10.5$; for $2Q=22$ two QE's each with $l_{\rm QE}=4$ and $l_h=11$; for $2Q=23$ two QE's each
with $l_{\rm QE}=4.5$ and $l_h=11.5$; and for $2Q=24$ no QE and $l_h=12$.
Adding the angular momenta of the identical Fermion QE's gives $L_e$, the electron angular momenta
of the lowest band; adding to $L_e$ the angular momentum $l_h$ gives the set of allowed multiplets
appearing in the low energy sector.
For example, in Fig.~\ref{fig5}($\rm b''$) the allowed values of $L_e$ are $1\oplus 3\oplus 5\oplus 
7$, and the
multiplets at 7 and 3 have lower energy than at 1 and 5.
Four low energy bands appear at $4\leq L \leq 18$, $8 \leq L \leq 14$, $6\leq L \leq 16$, and
$10 \leq L \leq 12$, resulting from $L_e=7$, 3, 5, and 1, respectively.

For $d \approx 1$, the electron--hole interaction results in formation of bound states of a hole and
one or more QE's.
In the two-electron--one-hole system, the $X$ and $X^-$ unbind for $d\approx 1$, but interaction
with the surrounding unbound electrons in a larger system can lead to persistence of these excitonic
states beyond $d=1$.
For example, the band of states at $d=0$ in Fig.~\ref{fig5}(a) that we associated with an $X^-$ 
interaction
with a QH persists at $d=1.5$ in Fig.~\ref{fig5}($\rm a'$).
However, it appears to cross another low energy band that extends from $L=3$ to 8.
This latter band can be interpreted in terms of three QE's interacting with the hole as was done in
the weak coupling limit shown in Fig.~\ref{fig5}($\rm a''$).
The other bands of the weak coupling regime (those beginning at $L=5$, 6, 7, 8, and 9) have
disappeared into the continuum of higher states as a result of the increase of $V_{eh}$.

For $2Q=22$, the lowest band can be interpreted in terms of one $X^-$ interacting with two QH's of
the generalized CF picture.
The $X^-$ has $l_{X^-}^*=3$ and the QH's each have $l_{\rm QH}=4$.
The allowed values of $L_{\rm 2QH}$ are 7, 5, 3, and 1, and the {\em molecular} state ${\rm QH}_2$
which has the smallest average QH--QH distance would have $l_{{\rm QH}_2}=7$.
This gives a band of $X^-+{\rm QH}_2$ states going from $L=l_{{\rm QH}_2}-l_{X^-}^*=4$ to
$L=l_{{\rm QH}_2}+l_{X^-}^*=10$.
A higher band beginning at $L=2$ might be associated with a 2QH state at $L_{\rm 2QH}=5$ interacting
with an $X^-$.
The origin of the other bands is less certain.

For $2Q=23$, there are two low lying bands.
The first contains a hole with $l_{\rm QE}=4.5$.
This gives rise to a band beginning at $L=7$ of which only the lowest two members are indicated.
A second band appears to contain an additional QE--QH pair.
The cost of energy in creating this additional pair is comparable to the energy gained through the
interaction of the additional QE with the hole.
The lowest $h {\rm QE}_2$ state occurs at $l_{h{\rm QE}_2}=l_h-l_{{\rm QE}_2}=3.5$ (this results 
from
choosing $l_{\rm 2QE}=8$, the largest value from the set of allowed $L_{2QE}=8$, 6, 4, 2, and 0)
and adding $l_{h{\rm QE}_2}$ to $l_{\rm QH}=3.5$ to obtain a band with $0 \leq L \leq 7$.
The state with $L=7$ is missing, probably due to the large QE--QH repulsion at $l_{\rm QE-QH}=1$.
The bands occurring at $2Q=25$ are even more uncertain.

\subsection{Photoluminescence Selection Rules at High Magnetic Fields}

Exact numerical diagonalization gives both the eigenvalues and the eigenfunctions.
The low energy states $\left|i\right>$ of the initial $N$-electron--one-hole system have just been 
discussed.
The final states $\left|f\right>$ contain $N'=N-1$ electrons and no holes.
The recombination of an electron--hole pair is proportional to the square of the matrix element of
the photoluminescence operator $\hat \mathcal{L}$, where 
$\hat \mathcal{L}=\int d^2r \Psi_e(\vec{r}) \Psi_h(\vec{r})$ and $\Psi_e$ (or $\Psi_h$) 
annihilates an electron (or hole).
We have evaluated $|<f|\hat \mathcal{L}\left|i\right>|^2$ for all of the low-lying initial states 
and have
found the following results \cite{prb63-1,prb63-2}.
(i) Conservation of the total angular momentum $L$ is at most weakly violated through the scattering
of {\em spectator} particles (electrons and quasiparticles) which do not participate directly in the
recombination process if the filling factor $\nu$ is less approximately 1/3.
(ii) In the strong coupling region, the neutral $X$ line is the dominant feature of the PL spectrum.
The $X^-{{\rm QH}_2}$ state has very small oscillator strength for radiative recombination.
(iii) For intermediate coupling, the $h{{\rm QE}_2}$ and an excited state of the $h{\rm QE}$ (which 
we
denote by $h{{\rm QE}^*}$) are the only states with large oscillator strength for photoluminescence.

At zero temperature ($T=0$), all initial states must be ground states of the $N$-electron--one-hole
system.
At finite but low temperatures, excited initial states contribute to the PL spectrum.
The photoluminescence intensity is proportional to
\begin{equation}
w_{i \rightarrow f}=\frac{2\pi}{\hbar} \mathcal{Z}^{-1} 
\sum_{i,f} e^{-\beta E_i}\mid\left<f|\hat \mathcal{L}|i\right>\mid^2 \delta(E_i-E_f-\hbar\omega),
\end{equation}
where $\beta=(kT)^{-1}$ and $\mathcal{Z}=\sum_i e^{-\beta E_i}$.

\subsection{Singlet and Triplet Charged Excitons at Low Magnetic Fields}

As mentioned in Section VIII B, only spin polarized charged excitons (with $J_e=1$) are bound when
the ratio $(\hbar \omega_c)/(e^2/\lambda)$ tends to infinity.
In real systems at finite values of this parameter, both singlet ($J_e=0$) and triplet ($J_e=1$)
charged excitons occur.
According to the theory \cite{whittaker} the singlet $X_s^-$ is the ground state (GS) at low values
of the magnetic field, while the triplet $X_t^-$ is the GS at very high magnetic fields.
Numerical calculations of the ground states of both the singlet and triplet charged excitons
\cite{whittaker} indicated a crossing at roughly 30 Tesla for a symmetric GaAs quantum well, the
width of which was about 10 nm.
Observation of PL by Hayne {\sl et al.} \cite{hayne} displaying three peaks that were interpreted as
the $X$, $X_t^-$, and $X_s^-$, showed no crossing of the $X_t^-$ and $X_s^-$ up to the fields of 50
Tesla.
This led the experimenters to question the validity of the variational calculations.

In this section we study very small systems (either two or three electrons and one valence band
hole) in narrow ($\sim$ 11.5 nm) symmetric GaAs quantum wells.
We include the effects of Landau level mixing caused by the interactions, and the effect of finite
well width on the effective interaction.
Only a single subband is used in the calculations, since the quantum well is relatively narrow.
Both electrons and holes are described in the effective mass approximation, and interband coupling
is partially accounted for by a magnetic field dependence of the cyclotron mass of the hole (taken
from experimental data) \cite{cole}.
The Zeeman energy depends on both the well width and the magnetic field $B$.
Five Landau levels for both the electrons and holes were included in the calculation in order to
obtain satisfactory convergence.
The energies obtained for different values of the monopole strength $2Q$ were extrapolated to the
large $Q$ limit to eliminate finite-size effects.

\begin{figure}
\resizebox{15.0cm}{9.0cm}{\includegraphics{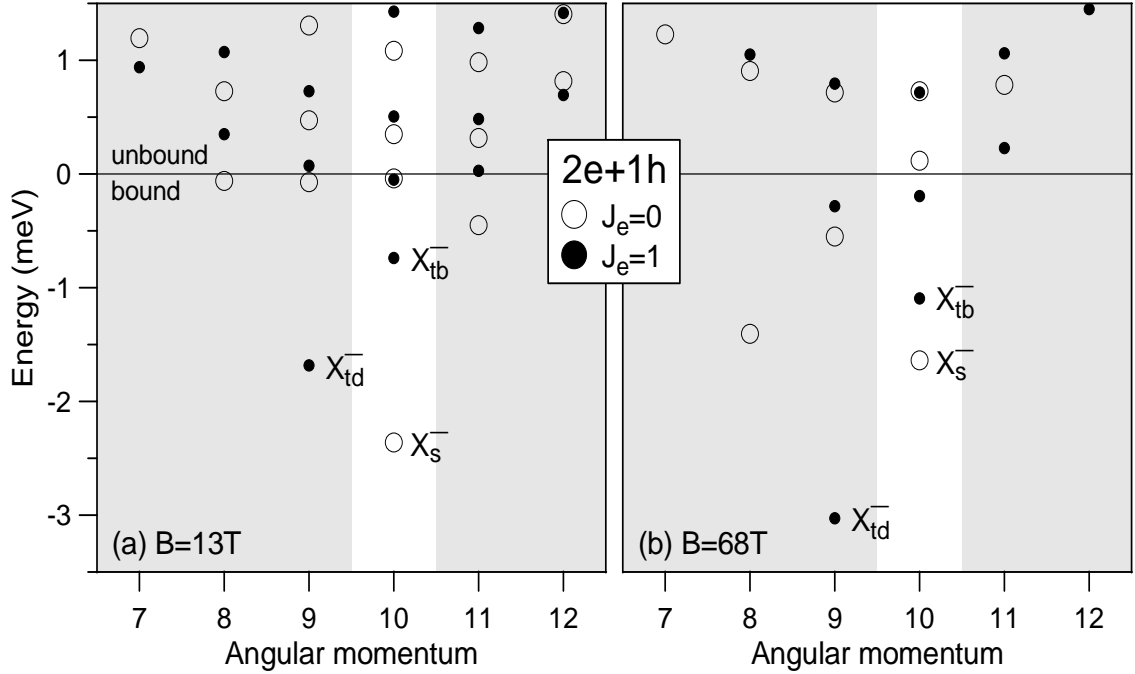}}
\caption{\label{fig6}
   Energy spectra (binding energy vs. angular momentum) of the two-electron--one-hole system 
   on a Haldane sphere with the Landau level degeneracy of $2Q+1=21$.
   $J_e$ denotes the total electron spin. 
   The parameters are appropriate for the 11.5 nm GaAs quantum well \cite{pe8}.}
\end{figure}
The energy spectra of the two-electron--one-hole system calculated for $2Q=20$ are shown 
in Fig.~\ref{fig6} \cite{pe8}.
Open and solid symbols mark singlet and triplet states ($J_e$ is the total electron spin), and each
state with $L>0$ represents a degenerate $L$ multiplet.
Since the PL process (annihilation of an $e$--$h$ pair and emission of a photon) occurs with
conservation of angular momentum, only states from the $L=Q$ channel are radiative 
\cite{pb256,prb60-2}.
Recombination of other non-radiative states requires breaking rotational symmetry (e.g., by 
collisions with electrons).
This result is {\em independent} of the chosen spherical geometry and holds also for a planar
quantum well, except that the definition of the conserved momentum is different \cite{pe3}.

The occurrence of a strict PL selection rule at finite $B$ may seem surprising, since the 
{\em hidden symmetry} \cite{lerner,dzyubenko} that forbids the $X_{td}^-$ recombination in the
lowest Landau level does not hold when the mixing with higher Landau levels is included.  
(The ``d'' in $X_{td}^-$ means ``dark'' and $X_{td}^-$ is called the {\em dark triplet} because it 
is
forbidden to decay radiatively.) 
However, it is both the hidden symmetry and the above-mentioned angular momentum conservation that
independently forbid the $X_{td}^-$ recombination, and the latter remains valid at finite $B$.
Although the hidden symmetry and resulting $N_X$ conservation law no longer hold at finite $B$, the 
$X_{td}^-$ recombination remains strictly forbidden because of the independently conserved $L$.

We expect breaking of both symmetries for real experimental situations.
The presence of impurities and defects, and $e$--$X_{td}^-$ scattering during recombination in the
presence of excess electrons can relax the strict conservation of the $X^-$ angular momentum in the
radiative decay.
However, for narrow and symmetric quantum wells containing a relatively small number of excess
electrons, the symmetries may be only weakly broken and some remnant of the strict conservation laws
may survive.

\begin{figure}
\resizebox{15.0cm}{9.0cm}{\includegraphics{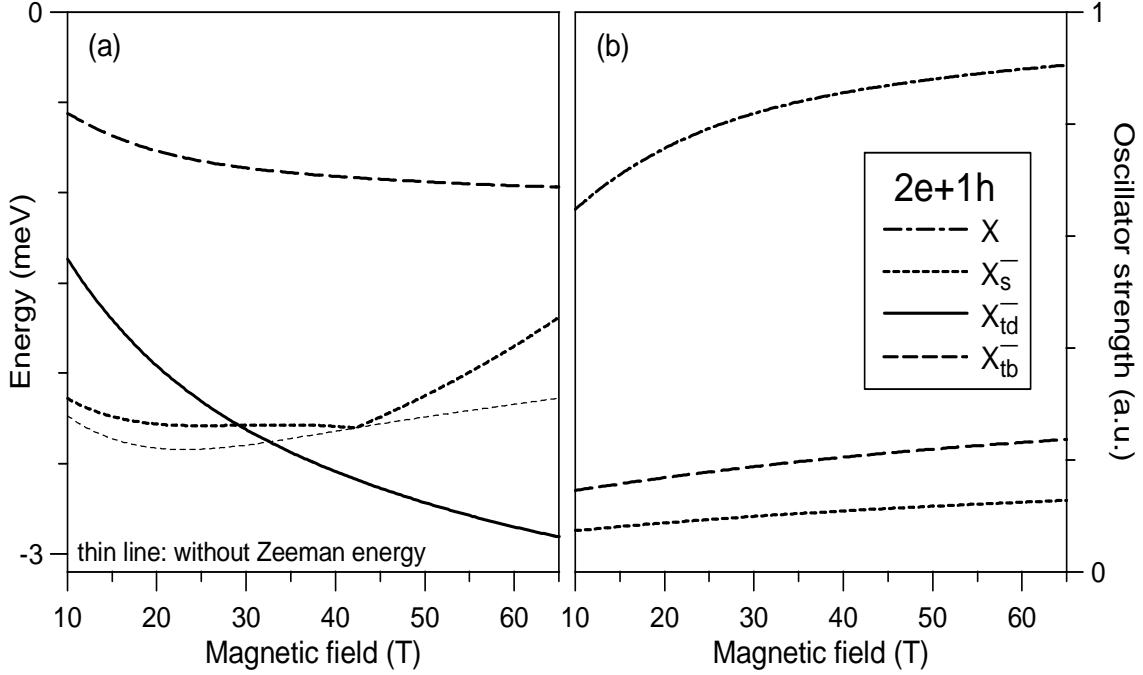}}
\caption{\label{fig7}
   The $X^-$ energies (a) and oscillator strengths (b) in the 11.5 nm GaAS quantum 
   well plotted as a function of the magnetic field \cite{pe8}.}
\end{figure}

Three states marked in Fig.~\ref{fig6} are of particular importance: $X_s^-$ and $X_{tb}^-$ (``b'' 
stands for
``bright'') are the only strongly bound radiative states, while $X_{td}^-$ has by far the lowest
energy of all non-radiative states.
The radiative triplet bound state $X_{tb}^-$ is identified for the first time.
The binding energies of all three $X^-$ states are extrapolated to $\lambda/R \rightarrow 0$ and
plotted in Fig.~\ref{fig7}(a) as a function of $B$ \cite{pe8}.
For the $X_s^-$, the binding energy differs from the PL energy (indicated by thin dotted line) by
the Zeeman energy needed to flip one electron's spin, and the cusp at $B \approx 42$ T is due to the
change of sign of the electron $g$-factor.
For the triplet states, the PL and binding energies are equal.
The energies of $X_s^-$ and $X_{td}^-$  behave as expected: The binding of $X_s^-$ weakens at higher
$B$ and eventually leads to its unbinding in the infinite field limit \cite{macdonald};
the binding energy of $X_{td}^-$  changes as $e^2/\lambda \propto \sqrt{B}$; and the
predicted \cite{whittaker} transition from the $X_s^-$ to the  $X_{td}^-$ GS at $B \approx 30$ T is
confirmed.
The new $X_{tb}^-$ state remains an excited triplet state at all values of $B$, and its binding
energy is smaller than that of $X_s^-$ by about 1.5 meV.
The oscillator strengths $\tau^{-1}$ of a neutral exciton $X$ and the two radiative $X^-$ states are
plotted in Fig.~\ref{fig7}(b).
In the two-electron--one hole spectrum, the strongly bound $X_s^-$ and $X_{tb}^-$ states share a
considerable fraction of the total oscillator strength of one $X$, with $\tau_{tb}^{-1}$ nearly
twice larger than $\tau_s^{-1}$.

The comparison of calculated magnitude and magnetic field dependence of the $X^-$ binding energies
with the experimental PL spectra \cite{hayne,shields,finkelstein1,finkelstein2}, as well as high
oscillator strength of the $X_{tb}^-$,
lead to the conclusion that the three peaks observed in PL are the $X$, $X_s^-$, and $X_{tb}^-$.

To understand why the $X_{td}^-$ state remains optically inactive even in the presence of
collisions, the $e$--$X^-$ interaction must be studied in greater detail.
Our numerical results for a three electron--one hole system indicate that the lowest band of states
consists of a triplet $X^-$ and one unbound electron.
Because the $X_t^--e$ pseudopotential is superharmonic, in real experimental systems at a low
electron concentration ($\nu \leq 1/3$) Laughlin correlations between the electron and $X_t^-$ will
effectively isolate the $X_t^-$ from the surrounding 2DES.
This prevents close collisions of the $X_t^-$ and the {\em spectator} electron during the $e$--$h$
recombination.
Although the $X_{td}^-$ is no longer forbidden to decay radiatively since the {\em spectator}
electron can change its angular momentum in the recombination process, this scattering process is
weak for $\nu<1/3$.
The oscillator strength for radiative decay of the $X_{td}^-$ is found to be more than an order of
magnitude smaller than those of the $X_s^-$ and $X_{tb}^-$.
These results support the interpretation that the three peaks observed in many experiments
correspond to the $X$, $X_s^-$, and $X_{tb}^-$.
The $X_{td}^-$ is not observed because of its small oscillator strength.
The $X_{td}^-$ recombination line has been observed only recently, when special care (very low
temperatures and high quality samples) were taken to detect its weak signal \cite{yusa}.
Even more convincing is the comparison with infrared absorption at very low temperature where only
the $X_{td}^-$ state is heavily occupied.
Absorption spectra show only one strong peak in contrast to PL spectra which shows three, because
the higher population of the $X_{td}^-$ compensates for its lower oscillator strength for radiative
recombination compared to the $X_s^-$ and $X_{tb}^-$.

\subsection{Skyrmion Excitons}

In order to understand the excitonic complexes that can be formed near filling factor $\nu=1$, it is
first necessary to study the kinds of elementary excitations than can occur in the absence of
valence band holes.
For filling factor $\nu$ equal to unity, the lowest energy excitations are spin flip excitations
which create a reversed spin electron $e_{\rm R}$ in the same $n=0$ Landau level leaving behind 
a spin hole $h$ in the otherwise filled $\nu=1$ state.
Even when the Zeeman energy $E_{\rm Z}$ is zero, the Coulomb exchange energy will spontaneously
break the spin ($\uparrow,\downarrow$) symmetry giving a spin polarized ground state.

\begin{figure}
\resizebox{15.0cm}{9.0cm}{\includegraphics{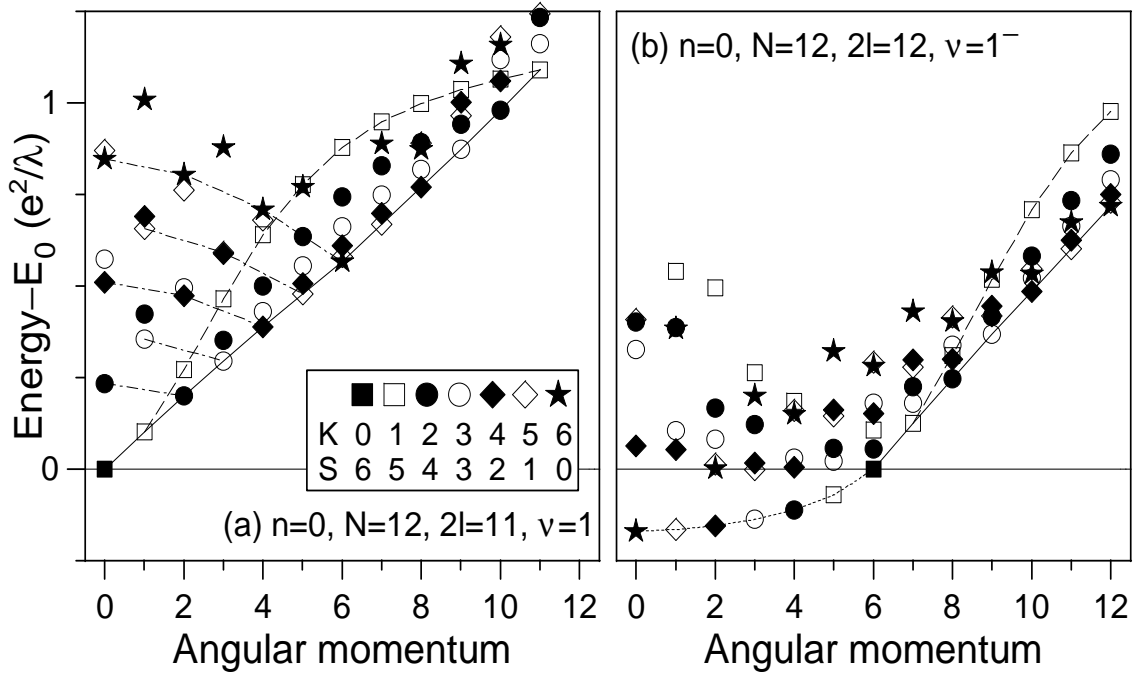}}
\caption{\label{fig8}
   Energy spectra (energy $E$ vs. angular momentum $L$) of the spin unpolarized 
   12-electron system in the lowest Landau level, calculated on a Haldane sphere 
   for monopole strength $2Q=11$ (a) and 12 (b) \cite{ssc122}.}
\end{figure}

In Fig.~\ref{fig8}(a) we show the low lying spin excitations of the $\nu=1$ state (with $E_{\rm Z}$ 
taken to
be zero) for a system of $N=12$ electrons \cite{ssc122}.
The solid square at $L=0$ is the spin polarized $\nu=1$ ground state with spin $S=6$.
The symbol $K=N/2-S$ is the number of spin flips away from the fully spin polarized state.
The band of open squares connected by a dashed line gives the spin wave (SW) dispersions
$\varepsilon_{\rm SW}(L)$.
The angular momentum $L$ is related to wavenumber $k$ by $L=kR$, where $R$ is the radius of the
spherical surface to which the $N$ electrons are confined.
The SW consists of a single $e_{\rm R}h$ pair; its dispersion can be evaluated
analytically \cite{kallin}.
The solid circles, open circles, etc. represent states containing 2, 3, $\dots$ spin flips
(i.e., 2, 3, $\dots e_{\rm R}h$ pairs).
Dot-dashed lines connect low lying states with equal numbers of spin flips.
It is interesting to note the almost straight line connecting the lowest energy states at $0 \leq
L \leq 6$.
This can be interpreted as band of $K$ SW's each with $l_{\rm SW}=1$ with $L=K$.
The near linearity suggests that these $K$ SW's are very nearly noninteracting in the state with
$L=K$.

In Fig.~\ref{fig8}(b) we present the lowest energy spectra for $\nu=1^-$ (i.e., a single spin hole 
in the
$\nu=1$ state).
In both Fig.~\ref{fig8}(a) and (b) only the lowest energy states at each $L$ and $S$ are shown.
Of particular interest in Fig.~\ref{fig8}(b) is the band of states with $L=S=Q-K$ and negative 
energy.
These are antiskyrmion states, $S_K^+=K\mbox{ }e_{\rm R}+(K+1)h$, bound states of one spin hole
and $K$ spin waves \cite{fertig,ssc122}.
They are analogous to interband charged excitons \cite{charged excitons,szlufarska}, but they can be 
equilibrium 
states not
subject to radiative decay at the appropriate value of the Zeeman energy.
Skyrmion states are $S_K^-=K\mbox{ }h+(K+1)e_{\rm R}$.
Electron--hole symmetry requires their existence for $\nu>1$.

It has been demonstrated \cite{ssc122} that in the fractional quantum Hall regime analogous
excitations occur with ${\rm QE}_{\rm R}$ and QH replacing $e_{\rm R}h$ and $h$ of the integral 
quantum Hall
case.
Spin waves, skyrmions, and antiskyrmions made from Laughlin quasiparticles occur for $\nu \approx 
1/3$.
The most stable skyrmion or antiskyrmion size depends weakly on the quantum well width for the $\nu 
\approx 1$ state, but for $\nu \approx 3, 5, \dots$ the well width $w$ must be of the 
order of a few times the magnetic length in order to obtain stable bound states of spin waves and 
spin holes or reversed spin electrons \cite{wqskyrmion,ssc122,cooper}.

The skyrmion and antiskyrmion states $S_K^{\pm}$ are quite analogous to the excitonic $X_K^{\pm}$ 
states of valence band holes interacting with conduction band electrons.
In the ideal theoretical model, a valence hole has exactly the same interaction as a spin hole in 
the $\nu=1$ state of the conduction band.
In fact  these two types of holes can probably be distinguished by an pseudospin as is done for 
electrons on different layers of a bilayer system \cite{bilayer,brey,marinescu}.
The spectrum and possible condensed states of a multicomponent Fermion liquid containing electrons, 
$X^-, X_2^-$, etc., has been considered in Ref.~\cite{prb60-1}.
Exactly the same ideas are applicable to a liquid of electrons and skyrmions or antiskyrmions of 
different sizes.
The only difference is that the skyrmion $S^-=h(e_{\rm R})_2$ is stable while the $X^-=ve_2$ has a 
finite lifetime for radiative recombination of $e$--$v$ pair.

When there are $N_h$ spin holes in the $\nu=1$ level (or $N_e$ reversed spin electrons in addition
to the filled $\nu=1$ level) and when $N_h$ (or $N_e$) is much smaller than $N \approx 2Q+1$, the
degeneracy of the filled lowest Landau level, then the most stable configuration will consist of
$N_h$ antiskyrmions (or $N_e$ skyrmions) of the most stable size.
These antiskyrmions (or skyrmions) repel one another.
They are positively (or negatively) charged Fermions with standard Landau level structure, so it is
not surprising that they would form either a Wigner lattice or a Laughlin condensed state with $\nu$
for the antiskyrmion (or skyrmion) equal to an odd denominator fraction as discussed in
Refs.~\cite{ssc122,prb60-2,skyrmions,cote,paredes,wqskyrmion}.

In the ideal theoretical model, a valence hole acts exactly like a spin hole in the $\nu=1$ level 
of the conduction band.
Therefore we would expect an excitonic complex consisting of $K$ spin waves bound to the valence
hole to be the lowest energy state, in the same way that the antiskyrmion consisting of $K$ spin
waves bound to a spin hole in the $\nu=1$ level gives the lowest energy state when $E_{\rm Z}$ is
sufficiently small.
For a small number of valence holes, the $X_K^+=v(e_{\rm R}h)_K$ excitonic complexes formed by
each valence hole will repel one another.
If a small number of antiskyrmions are already present (for $\nu<1$), the positively charged
antiskyrmion--charged exciton repulsion will lead to Laughlin correlations or Wigner crystallization
of the multicomponent Fermion liquid.
Just as for the charged excitons ($X^-$) in the dilute regime, the PL at low temperature will be 
dominated by
the $X_K^+\rightarrow S_{K'}^++\gamma$ process, with $K'=K$ or $K-1$ depending on spin of the
annihilated valence hole (i.e., on the circular polarization of the emitted photon $\gamma$).
This corresponds to the most stable $X_K^+$ undergoing radiative $ev$ or $e_{\rm R}v$ recombination
and leaving behind an antiskyrmion consisting of $K$ or $K-1$ spin waves bound to a spin hole of the
$\nu=1$ state.
Because the valence hole and the spin hole in the $\nu=1$ conduction level are distinguishable (or
have different pseudospin) even in the ideal theoretical model this PL is not forbidden.
It will be very interesting to see how realistic sample effects (finite well width, Landau level
admixture, finite separation between the electron and valence hole layers) alter the conclusions of
the ideal theoretical model.

For $\nu \geq 1$, negatively charged skyrmions are present before the introduction of the valence
holes.
The skyrmions are attracted by the $X_K^+$ charged exciton, but how this interaction affects the PL
can only be guessed.
It is possible that the interaction of the valence hole with the skyrmions will lead to the
formation of an $X$ or an $X_{td}^-$ and spin waves.
The $X_{td}^-$ will be very weakly radiative (just as in the case of $\nu \leq 1/3$).
However, the recombination can occur with a majority spin electron.
This case was considered in Ref.~\cite{neg skyrmions,portengen} for the case of a single $X_{td}^-$.

We believe that numerical diagonalization for realistic models including Landau level admixture and
finite well width should explain the behavior of PL for electron filling factors close to unity.
Only qualitative behavior expected has been discussed in this work.
Realistic ``numerical experiments'' are being carried out to check whether the expected behavior is
correct.

\section{Summary}

This review contains a discussion of the energy spectra of a system of $N_e$ electrons and $N_h$
holes.
In photoluminescence, $e$--$h$ recombination leads to peaks in the spectra that depend on the 
initial
and final states.
For ideal systems, with $\hbar \omega_c \gg e^2/\lambda$, very narrow quantum wells, and 
$|V_{ee}|=|V_{eh}|$, the spectra contain information only about the neutral exciton.
For non-ideal systems, where $|V_{eh}|\ll|V_{ee}|$, spectra contain information about the
electron--electron correlations in the underlying electron gas.

The authors gratefully acknowledge the support by Grant DE-FG 02-97ER45657 of 
the Material Research Program of Basic Energy Sciences--US Department of Energy.
AW acknowledges support from Grant 2P03B02424 of the Polish KBN 
and KSY acknowledges partial supports by the ABRL(R14-2002-029-01002-0) through the KOSEF 
and by PNU(2003).

\end{document}